\documentclass[journal=nalefd,manuscript=letter,layout=twocolumn]{achemso}

\usepackage{chemformula}
\usepackage[T1]{fontenc}

\usepackage{graphicx}% Include figure files
\usepackage{dcolumn}% Align table columns on decimal point
\usepackage{bm}% bold math
\usepackage[mathlines]{lineno}% Enable numbering of text and display math
\usepackage{xcolor}

\usepackage[utf8]{inputenc}
\usepackage[T1]{fontenc}
\usepackage{mathptmx}
\usepackage{etoolbox}
\usepackage[version=4]{mhchem}
\usepackage{gensymb}
\usepackage{miller}
\usepackage{float}

\renewcommand{\selectlanguage}[1]{}

\makeatletter
\def\@email#1#2{%
 \endgroup
 \patchcmd{\titleblock@produce}
  {\frontmatter@RRAPformat}
  {\frontmatter@RRAPformat{\produce@RRAP{*#1\href{mailto:#2}{#2}}}\frontmatter@RRAPformat}
  {}{}
}%

\newcommand*{\citen}[1]{%
  \begingroup
    \romannumeral-`\x % remove space at the beginning of \setcitestyle
    \setcitestyle{numbers}%
    \cite{#1}%
  \endgroup   
}

\author{Michael Xu}
\author{Kevin Ye}
\author{Ida Sadeghi}
\author{R. Jaramillo}
\author{James M. LeBeau}
\affiliation{ 
Department of Materials Science and Engineering, Massachusetts Institute of Technology, Cambridge, MA 02139, USA}
\email{lebeau@mit.edu}

\title[]
  {Atomic Structure of Self-Buffered \ce{BaZr(S,Se)3} Epitaxial Thin Film Interfaces}


\begin{document}

\begin{abstract}

Understanding and controlling the growth of chalcogenide perovskite thin films through interface design is important for tailoring film properties. Here, the film and interface structure of \ce{BaZr(S,Se)3} thin films grown on \ce{LaAlO3} by molecular beam epitaxy and post-growth anion exchange is resolved using aberration-corrected scanning transmission electron microscopy. Epitaxial films are achieved from self-assembly of an interface ``buffer'' layer, which accommodates the large film/substrate lattice mismatch of nearly 40\% for the alloy film studied here. The self-assembled buffer layer, occurring for both the as-grown sulfide and post-selenization alloy films, is shown to have rock-salt-like atomic stacking akin to a Ruddlesden-Popper phase. Above this buffer, the film quickly transitions to the perovskite structure. Overall, these results provide insights into oxide-chalcogenide heteroepitaxial film growth, illustrating a process that yields relaxed, crystalline, epitaxial chalcogenide perovskite films that support ongoing studies of optoelectronic and device properties.

\end{abstract}

\maketitle

\section*{Main Text}

Growth of high-quality semiconductor thin films, needed for applications in microelectronics and optoelectonics, requires the availability of substrates that are well-matched structurally and chemically. This presents problems for many emerging materials for which suitable substrates are not readily available. For instance, many  chalcogenide semiconductors have chemistry, crystal structure, and lattice constants that are substantially different than available oxide and compound semiconductor substrates. Further, many proposed device concepts - including two-dimensional materials for transistors and chalcogenide perovskites for solar cells - may depend on better control over oxide-chalcogenide hetero-interfaces. Better control hinges on a better understanding of interfaces with many elements (here, La, Al, O, Ba, Zr, S, and Se) in atomic structures that may deviate substantially from nominal bulk phases and that are difficult to predict \textit{a priori}. Advancing semiconductor heteroepitaxial growth of new materials, therefore, depends on overcoming substantial challenges in advanced characterization. Here, we take on this challenge in the context of the growth of chalcogenide perovskite thin films on complex oxide substrates by molecular beam epitaxy (MBE). 

Chalcogenide perovskites of type \ce{ABX3} (\ce{A=Ca, Sr, Ba, B=Ti, Zr, Hf, X=S,Se}), with a direct band gap that is tunable within visible to near-infrared (VIS-NIR), strong optical absorption, low toxicity, and excellent environmental stability, are of interest for future solar cell technologies.\cite{Sun2015-sj, Niu2017-vj, Jaramillo2019-du, Nishigaki2020-hd} The most-studied chalcogenide perovskite, \ce{BaZrS3}, features strong dielectric response, strong optical absorption, exceptional thermal stability, and has been made as powders, crystals, nanocrystals, and thin films.\cite{Meng2016-sf, Niu2017-vj, Niu2018-wa, Wei2020-md,Filippone2020-dt, Nishigaki2020-hd, Zilevu2022-ul, Yang2022-kk, Ye2022-jw, Sadeghi2021-il} Epitaxial thin films in the perovskite structure have been grown by the authors across the full \ce{BaZrS_{3-y}Se_{y}, y=0-3} alloy range, with direct band gap spanning the range 1.5 - 1.9 eV.\cite{Sadeghi2021-il, Sadeghi2023-wv, Ye2024-va} Notably, despite a very large lattice constant mismatch of \ce{BaZrS3} with the \ce{LaAlO3} substrates, the formation of relaxed, epitaxial \ce{BaZrS3} films was observed to be accommodated by a self-assembled buffer layer at the interface.\cite{Sadeghi2021-il}

By comparing the bulk pseudocubic lattice parameters for the \ce{BaZrS3} film (497.5 pm) and \ce{LaAlO3} substrate (381.1 pm), a coincident site lattice (CSL) was proposed to explain the observed epitaxial relationship.\cite{Sadeghi2021-il} The appearance of structural perturbations at the interface and the potential for surface passivation and reconstruction in the presence of \ce{H2S} at high temperature, however, indicate that the interface is too complex to be described only by a CSL.\cite{Sadeghi2021-il, Klenov2005-ah, Wang2023-fk, Liu2017-rm} More recently, the authors realized thin film sulfide-selenide perovskite alloys by both direct growth and post-growth selenization methods.\cite{Sadeghi2023-wv, Ye2024-va} Beyond modifying the bandgap, the higher atomic mass of selenium compared to sulfur also enables a more detailed atomic-scale study of the interface. 

Here, the epitaxial interface between \ce{BaZrS3} and \ce{BaZr(S,Se)3} films with \ce{LaAlO3} oxide substrates is characterized using aberration-corrected scanning transmission electron microscopy (STEM). Post-growth selenization of \ce{BaZrS3} films through annealing in \ce{H2Se} gas, used to make the \ce{BaZr(S,Se)3} films, exchanges a fraction of the chalcogen ions without changing the interface structure or film microstructure. The self-assembled buffer layer is shown to have a rock-salt-like structure, similar to stacking faults often found in perovskites with imperfect cation stoichiometry and in Ruddlesden-Popper (RP) layered structures. This buffer layer accommodates the film-substrate mismatch and explains the structural deviations at and near the interface, which quickly diminish away from the interface.

\begin{figure}[!ht]
\centering
  \includegraphics[width=3.3in]{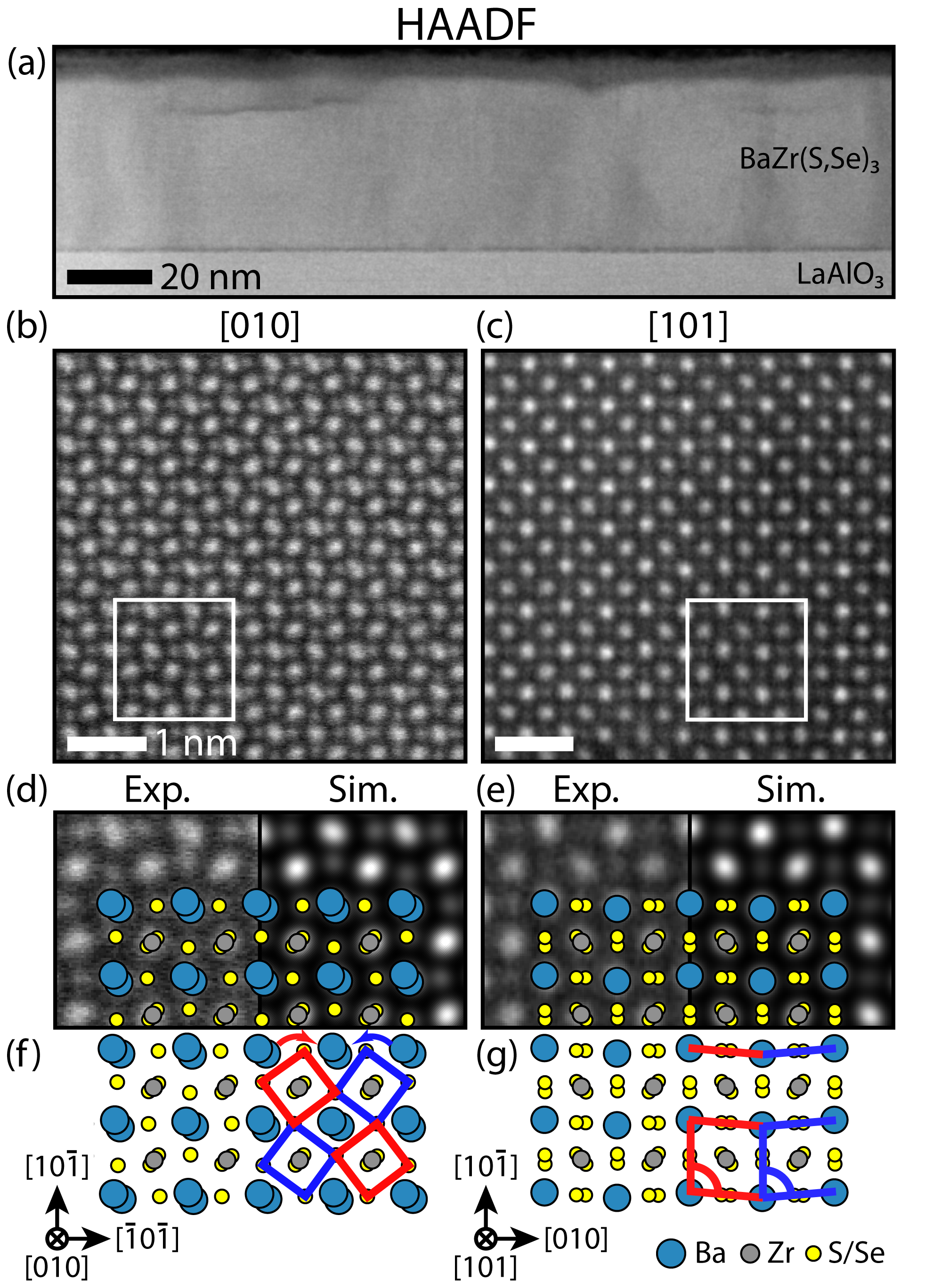}%
\caption{(a) STEM HAADF overview of the selenized thin film. Atomic resolution HAADF images of the (b) \hkl[010] and (c) \hkl[101] rotation variants. Comparison between experiment and STEM image simulations of the \textit{Pnma} (d) \hkl[010] and (e) \hkl[101] projections for \ce{BaZrS_{3-y}Se_{y}, y=2.21}. The corresponding projections of the structural models are overlaid for clarity, with (f) octahedral tilt angle and (g) Ba-Ba displacement angle illustrated. }%
\label{fig:fig1} 
\end{figure}

A high-angle annular dark-field (HAADF) STEM image of a 40-nm-thick selenized film is shown in Figure~\ref{fig:fig1}a. As previously reported for the pure sulfide, a self-assembled buffer layer and a ``cube-on-cube'' epitaxial relationship are seen at the interface.\cite{Sadeghi2021-il} Elemental quantification with energy dispersive X-ray spectroscopy (EDS) indicates a uniform distribution of Ba, Zr, S, and Se throughout the film cross-section, with an overall composition of \ce{BaZrS_{3-y}Se_{y}, y=2.21}.\cite{Ye2024-va} Similar to the pure sulfide \ce{BaZrS3} films, the selenized film is composed of 90\degree{} rotation variants, which originate due to the nearly-equivalent in-plane pseudocubic lattice parameters along the orthorhombic \hkl[010] and \hkl[101] projections.\cite{Sadeghi2021-il} These two domain types or rotation variants are shown in STEM HAADF images in Figure~\ref{fig:fig1}b,c and compared to STEM image simulations Figure~\ref{fig:fig1}d,e  assuming the orthorhombic structure with randomly-distributed anion occupancy according to the $y=2.21$ average composition. Experiment and simulation are in agreement, where tilting of the \ce{S}/\ce{Se} octahedra and alternating Ba-Ba displacements are visible in the \hkl[010] and \hkl[101] projected variants, respectively (Figure~\ref{fig:fig1}f,g). 

%Se ~ 40
% S ~ 18
% Ba ~ 20
% Zr ~ 22

\begin{figure*}[!ht]
\centering
  \includegraphics[width=6.5in]{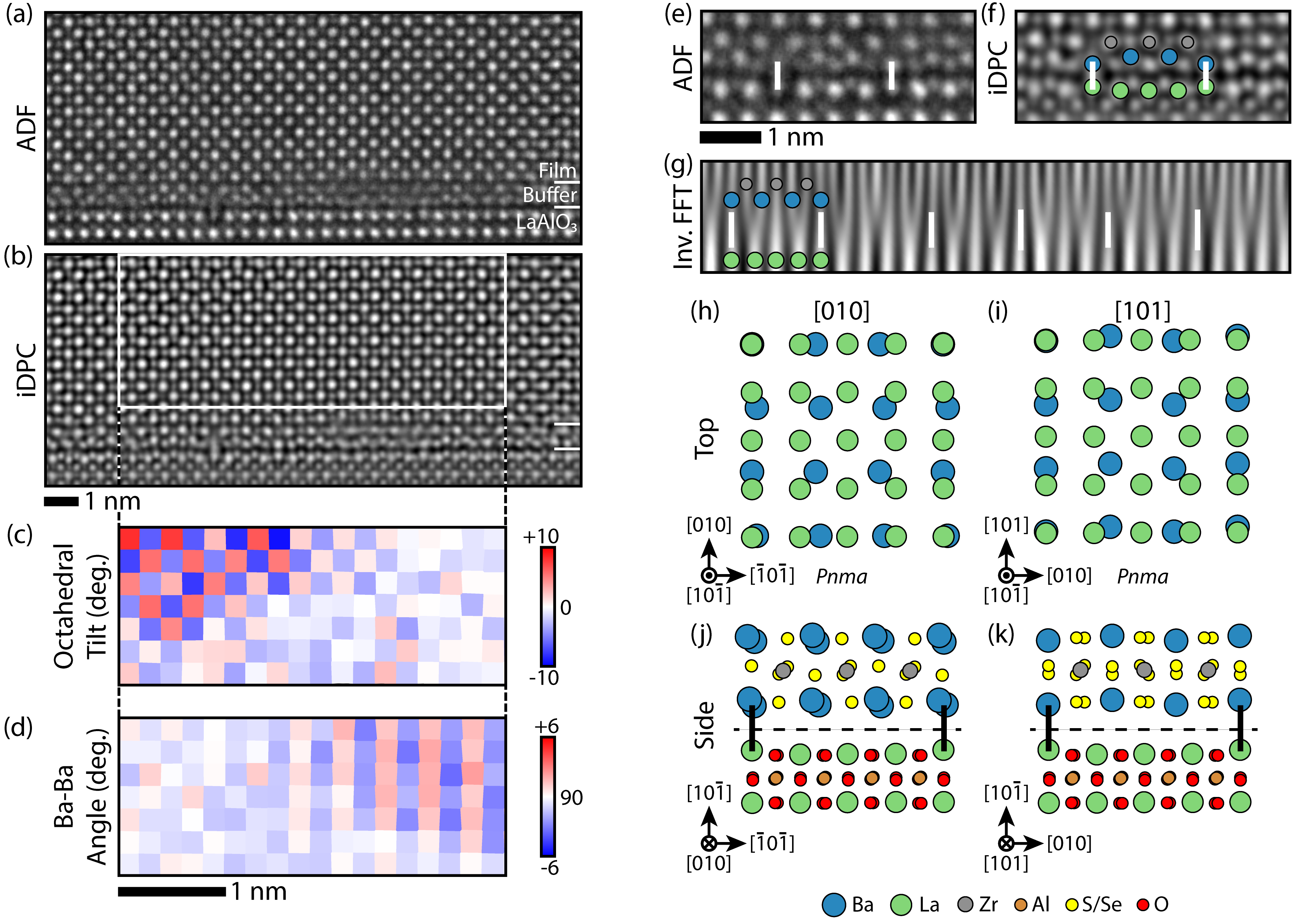}%
\caption{Simultaneously acquired (a) ADF and (b) iDPC images of the film interface. (c) Octahedral tilt angle for the \hkl[010] variant, and (d) Ba-Ba displacement angle for the\hkl[101] variant, for the boxed region in (b). Each square in the map represents one pseudocubic unit cell. (e) ADF and (f) iDPC images of the interface, with the coincident \ce{Ba} and \ce{La} sites connected by white lines. (g) inverse Fourier-filtered ADF image of the interface from (a) showing the alignment of the lattice planes and direct bonding sites (boxed). (h,i) Top and (j,k) side views of the interface models for the \hkl[010] and \hkl[101] rotation variants without coincident site displacements. The coincident sites are marked with connecting black lines.}%
\label{fig:fig2} 
\end{figure*}

The atomic structure of the film/substrate interface is highlighted in the simultaneously acquired annular dark-field (ADF) and integrated differential phase contrast (iDPC) images (Figure~\ref{fig:fig2}a,b). Approaching the interface, the anion octahedral tilt angle (Figure~\ref{fig:fig2}c) and Ba-Ba displacement angle (Figure~\ref{fig:fig2}d) are reduced compared to their values measured from the film bulk (6.4\degree{} and 3.0\degree{}, respectively). These changes near the interface are reminiscent of previous reports of perovskite oxide heteroepitaxy, where oxygen octahedra tilting is either dampened or propagated in the thin film through boundary constraints with the substrate.\cite{Liao2017-nk, Fowlie2019-hc, Jeong2020-eq, Lin2021-kw, Mun2023-at} 

% pure \ce{BaZrS3} (7.5\degree{} and 4.1\degree{}, respectively), consistent with the structural changes due to anion-exchange \cite{Xu2023-ne, Bak2019-ij}.
%the values for the ideal structure measured from simulated STEM images (7.5\degree{} and 4.1\degree{} for the \hkl[010] octahedral tilt and \hkl[101] Ba-Ba angles, respectively). 
%This gradual transition between rotation variants is likely due to the anion exchange of \ce{S}/\ce{Se}, which induces distortions to the lattice and \ce{Zr(S,Se)6} octahedra \cite{Xu2023-ne, Bak2019-ij}. 
% Rather than a planar boundary separating the rotation variants, as seen in \ce{BaZrS3}, a diffuse transition is present, in which anion octahedra tilt angles decrease in amplitude from left to right (Figure~\ref{fig:fig2}c). Concomitantly, the magnitude of oscillations in the in-plane Ba-Ba angle increases into the \hkl[101] variant (Figure~\ref{fig:fig2}d).

The first two atomic planes of the selenized film - immediately at the interface with \ce{LaAlO3} - exhibit structure contrasting with the rest of the film (Figure~\ref{fig:fig2}a,e). This observation is similar to that previously reported for the pure sulfide \ce{BaZrS3}, suggesting that selenization changes the chalcogen composition without affecting the interface structure. ADF, iDPC, and inverse Fourier filtered images (Figure~\ref{fig:fig2}e-g) illustrate an approximate coincidence of every five \ce{LaAlO3} unit cells with every four \ce{BaZr(S,Se)3} unit cells, in agreement with the overall periodicity observed from nano-beam electron diffraction (Supplementary Materials). At the coincident sites (connected by white lines in Figure~\ref{fig:fig2}e,f), the Ba and La atom columns are displaced closer to one another, suggesting stronger bonding at these locations. Based on this interface model, the top and cross-sectional views for the two 90\degree{} rotation variants are shown in Figure~\ref{fig:fig2}h-i, with in-plane lattice parameters determined from experiment (Supplementary Materials). The corners of the projected $5\times5$ square La sub-lattice are close to the $4\times4$ near-square Ba sub-lattice, exhibiting a mismatch of 0.5\% along \hkl[10-1] and 0.4\% along \hkl[010].

\begin{figure}[!h]
\centering
  \includegraphics[width=3.3in]{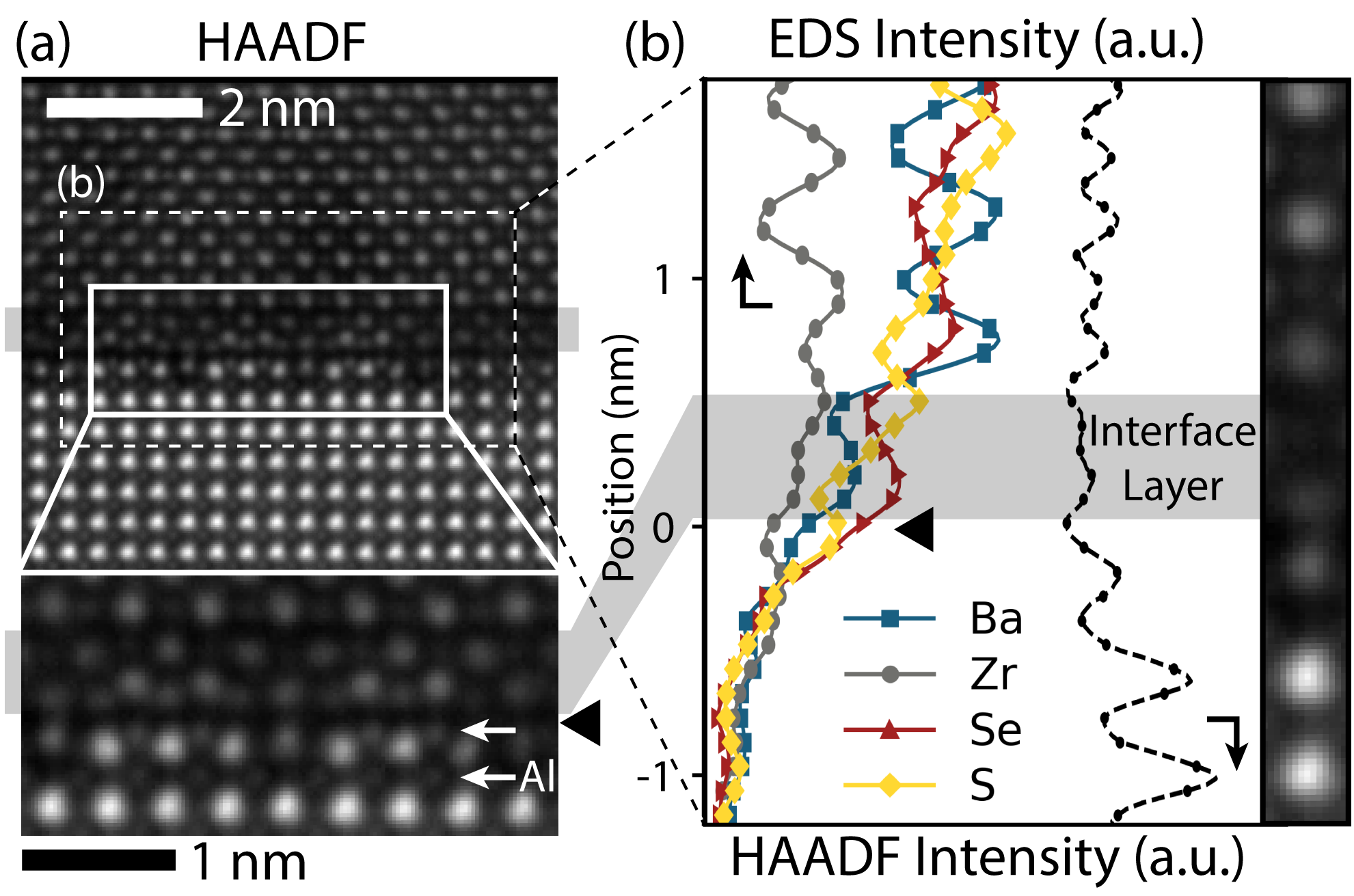}%
\caption{(a) HAADF overview of the film and substrate and inset of the buffer layer interface structure. Arrows mark the atom column in the gap and of \ce{Al} in the substrate. (b) Integrated line profile from the dashed boxed region in (a) showing EDS and HAADF intensity across the buffer layers and interface. The position of the adsorbed atom columns on the \ce{LaAlO3} surface is given by the black triangles.}%
\label{fig:fig3} 
\end{figure}

Despite this, the asymmetry of Ba positions in the orthorhombic structure yields imperfect coincidence for both of the 90\degree{} rotation variants. This is visible by the misalignment of Ba and La sub-lattices and incompatibility of both structures with the direct bonding coincident sites seen in experiment Figure~\ref{fig:fig2}h-k. From the structural model, the displacements of Ba are symmetric and opposite with respect to the La sub-lattice of the substrate, resulting in an additional mismatch of $\approx\pm$0.3\AA{}, yet in experiment, these displacements in addition to anion octahedral tilting are absent near the interface. In addition, from comparing images from experiment with the structural models (Figure~\ref{fig:fig2}e,f versus Figure~\ref{fig:fig2}j,k), additional atom columns are found within the interface gap. These atom column positions exhibit significant displacements compared to the bulk structures of both the \ce{BaZr(S,Se)3} film and \ce{LaAlO3} substrate. From atomic number-sensitive HAADF images of the interface (Figure~\ref{fig:fig3}a), the contrast of the atom columns found within the interface gap suggest occupation by \ce{S} ($Z=16$) or \ce{Se} ($Z=34$). This is evidenced by significantly brighter contrast than that of \ce{Al} ($Z=13$) columns in the \ce{LaAlO3} substrate (both marked by the white arrows) and lower contrast than those of \ce{Ba} ($Z=56$), \ce{La} ($Z=57$), or \ce{Zr} ($Z=40$). Further, the substrate surface is terminated by this plane of atom columns, suggesting that these atom columns and the rest of the buffer layer self-assemble during \ce{H2S} exposure of the substrate immediately before and during film growth.

% \ce{BaZrS3} 14.97195 1.1\%
% \ce{BaZrS3} 15.022 0.8\% \hkl[10-1]
% \ce{BaZrS3} 15.128 0.1\% \hkl[010]
% 15.01682671206
% \ce{Ba3Zr2S7} 15.052 0.7\%
% \ce{Ba3Zr2(S_{1/3}Se_{2/3})7} 0.1\%
% \ce{LaAlO3} 15.143 vs 15.189
% 15.108 .2% (.5%) +- .3 in either dxn, so a lot
% 15.116 .1% (.4%)

The self-assembled buffer layer is further characterized using EDS elemental mapping at the interface (Figure~\ref{fig:fig3}a,b and Supplementary Materials). The profile across the interface, shown in Figure~\ref{fig:fig3}b, is integrated along the plane of the interface to improve the signal-to-noise ratio with a comparatively reduced dose. A well-defined film is observed, with negligible diffusion beyond 1-2 nm across the film/substrate interface. Further, the HAADF image intensity is decreased within a 0.5 nm wide region at the interface (shaded region in Figure~\ref{fig:fig3}b) compared to the substrate and the rest of the film. Likewise, reduced HAADF intensity is observed at the terminating plane of the substrate immediately below the film, which can be explained by vacancies and/or static displacements of \ce{La} along the atom columns. Immediately adjacent to this plane of La atoms is a notable peak in the sulfur signal (triangle marker Figure~\ref{fig:fig3}a,b) whose location coincides with the unidentified atom columns extending into the interface gap (indicated with arrows in Figure~\ref{fig:fig3}a) These observations suggest that \ce{H2S} exposure at high temperature produces a subtle reconstruction of the \ce{LaAlO3} surface, including La-S bond formation.  

\begin{figure*}[!h]
\centering
  \includegraphics[width=6.2in]{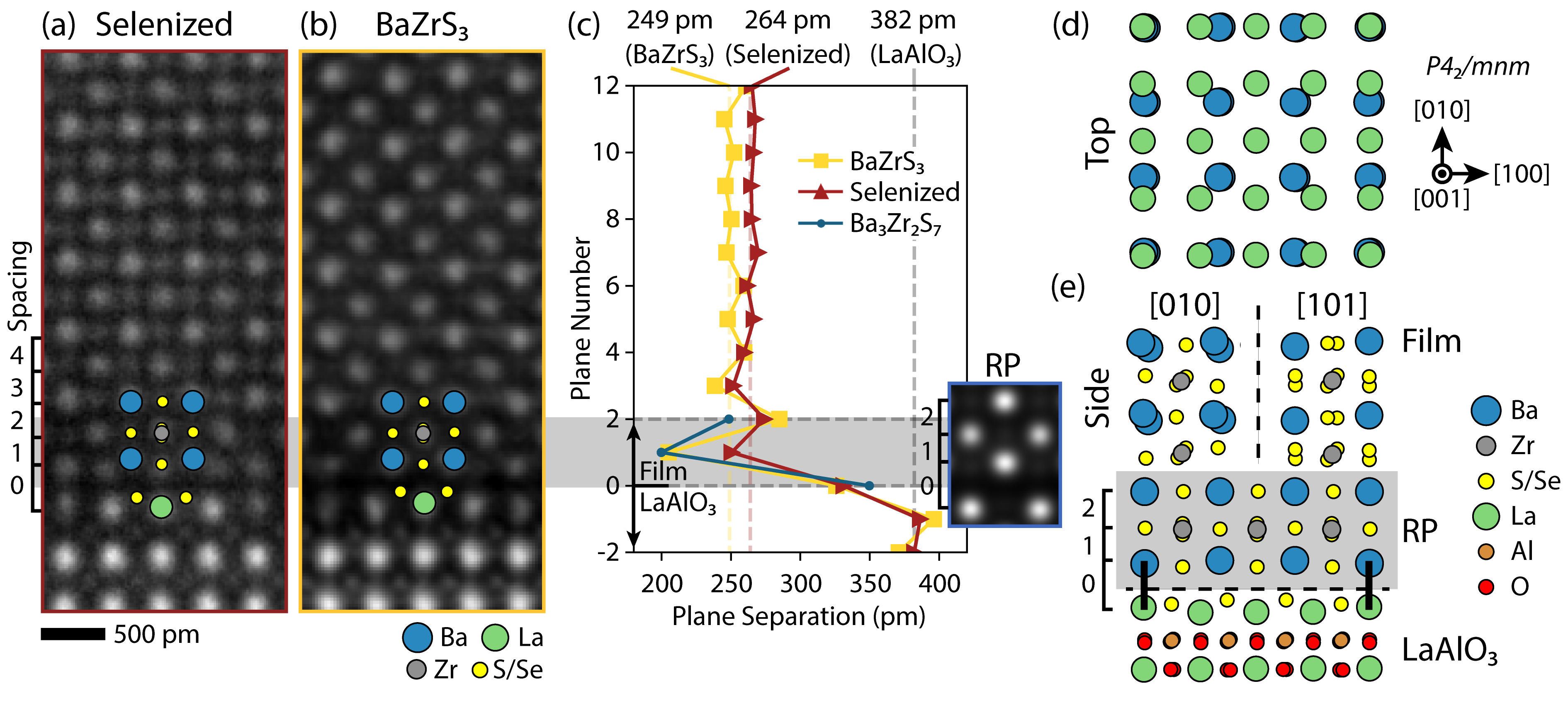}%
\caption{Interface HAADF images and overlaid RP models for the (a) selenized film and (b) \ce{BaZrS3} film. (c) Plane-to-plane distance calculated from the full images of (a-c) across the interface, with numbered spacings beginning at \#0, the film/substrate interface. (d) Side and (e) top views of the observed interface structure consisting of the \ce{LaAlO3} substrate, the RP-like buffer, and the orthorhombic perovskite.}
\label{fig:fig4}
\end{figure*}

The anion octahedral distortions and cation displacements in the self-assembled buffer layer are similar to the structure of the \ce{Ba3Zr2S7} RP phase (space group \textit{$P4_{2}mnm$}), which is shown in Figure~\ref{fig:fig4}a,b overlaid on HAADF images acquired on selenized and pure-sulfide films.\cite{Chen1994-rq, Niu2019-et} To highlight similarities in structure, interplanar distances, calculated from the in-plane averaged atom column positions, are measured from the acquired images of the grown films and simulated images of the RP phase and shown in Figure~\ref{fig:fig4}c. These spacings are numbered beginning at the film/substrate interface (\#0). In the first plane of the films, the anion atom columns are displaced towards the interface gap (Figure~\ref{fig:fig4}a,b). This displacement is characteristic of the rock-salt layers in the RP phase, in which anion planes terminate each perovskite slab and extend into the gap, and connects the structure of the perovskite film to that of the passivated substrate.

Further quantifying the interplanar distances within the buffer layer(Figure~\ref{fig:fig4}c) shows agreement with the RP structure. The average separation of 380 pm in the substrate is consistent with the pseudocubic lattice parameter of \ce{LaAlO3}. Across the interface, or spacing \#0, the distances between \ce{La} and \ce{Ba} planes in the selenized and pure sulfide films are 331 pm and 325 pm, respectively, close to the ideal rock-salt layer separation of 350 pm in \ce{Ba3Zr2S7} (Figure~\ref{fig:fig4}c). The next interplanar distances (spacing \#1) are 250 pm in the selenized and 205 pm in the pure sulfide films. This contraction is consistent with the RP structure, in which the planar spacing immediately adjacent to the rock-salt later is contracted (to 200 pm) due to the polar displacements of Ba and the \ce{ZrS6} octahedra.\cite{Wang2016-rk} A subsequent increase in interplanar distance at the Ba-\ce{ZrS6} octahedra (spacing \#2) is again similar to the RP structure. Beyond spacing \#2 we observe a reduction in the oscillation of planar separations, converging to an average out-of-plane interplanar distance of 251 pm for \ce{BaZrS3} and 266 pm for the selenized film, matching the expected perovskite structure (\textit{Pnma}).\cite{Niu2019-et, Ye2024-va} Patterns of anion octahedral tilts and Ba-Ba angle variations (Figure~\ref{fig:fig1}e), absent in the RP structure, emerge and become correlated within 1-3 unit cells of the buffer layer, similar to observations at perovskite oxide interfaces.\cite{Stone2016-za, Mun2023-at} Finally, even with anion exchange of \ce{S} with \ce{Se}, the structure of the RP buffer is preserved, with aligned Ba and La coincident planes at the interface. 

In conclusion, the formation of epitaxial and relaxed \ce{BaZrS3} and \ce{BaZr(S,Se)3} chalcogenide perovskite films on \ce{LaAlO3} substrates with large lattice mismatch is made possible by a self-assembled buffer layer. This layer likely forms during the initial stages of \ce{BaZrS3} film growth, including annealing and \ce{H2S} gas exposure of the substrate. The buffer layer resembles the rock-salt-like layers found in RP phases, and the interface structure is quantitatively comparable to \ce{Ba3Zr2S7}. Notably, the selenization process, used to convert the pure sulfide to an alloy film, appears to substitute Se for S throughout the film, including in the buffer layer, without affecting the buffer layer structure or its strain relief functionality. The epitaxial growth and facile selenization process associated with this buffer layer enables study of the optoelectronic properties of chalcogenide perovskite alloys with a wide range of composition, band gap, and lattice constant, for which matching substrates may not be available. Finally, the gas-source film growth method used here may be more broadly applicable to heteroepitaxial growth of chalcogenide semiconductors on non-matched, non-chalcogenide substrates, if the self-assembled buffer layer is found to generalize beyond the \ce{LaAlO3}/\ce{BaZr(S,Se)3} system.

% which has a tetragonal structure ($P4_2mnm$) 
% i4mmm\cite{Chen1994-rq}
% fmmm \cite{Chen1993-zy}
% p42mnm \cite{Niu2019-et}

\section*{Methods}
\subsection*{Film Growth}
Thin films were grown and annealed using a chalcogenide molecular beam epitaxy (MBE) system (Mantis Deposition M500). Further details on the deposition system and protocols are described in Ref~\citen{Ye2024-va}. Thin films were grown on 10 x 10 x 0.5 mm$^3$ (001)$_{PC}$-oriented \ce{LaAlO3} substrates (MTI). PC stands for pseudo-cubic; (001)$_{PC}$ corresponds to the (012) family of reflections in a rhombohedral system. Prior to growth, the substrates were outgassed in the MBE chamber at 1000\degree{} C under \ce{H2S} gas. Epitaxial \ce{BaZrS3} films were grown at 900-1000\degree{} C with elemental Zr and Ba sources, under 0.8 sccm \ce{H2S} flow. \ce{BaZrS3} epitaxial thin films were alloyed with selenium by annealing them under a mixture of 0.2 sccm \ce{H2S} and 0.2 sccm \ce{H2Se} at 800\degree{} C for 60 minutes in the MBE chamber. 

\subsection*{STEM Characterization}
Cross-sectional samples for electron microscopy were prepared by nonaqueous mechanical wedge polishing followed by single sector \ce{Ar+} ion milling (Fischione 1051 TEM Mill).\cite{Dieterle2011-da} A probe aberration-corrected Thermo Fisher Scientific Themis Z S/TEM operated at 200 kV was used for STEM imaging and spectroscopy. Imaging was performed using a probe semi-convergence angle of 18 mrad and probe current of 30 pA. High-angle annular dark-field (HAADF), annular dark-field (ADF), and integrated differential phase contrast (iDPC) images were acquired using collection angles of 65-200 mrad, 25-153 mrad and 6-24 mrad, respectively, and a dwell time of 1~{\textmu}s. The revolving STEM (revSTEM) method was used to correct for image distortions from sample drift during imaging for the 12-frame $1024\times1024$ image series.\cite{Sang2014-wz} STEM image simulations were performed using a custom implementation of the multislice method.\cite{Kirkland2010-vu} STEM energy dispersive X-ray spectroscopy (EDS) maps were acquired using SuperX detectors and quantified using the Thermo Fisher Scientific Velox software. A 5 pixel averaging pre-filter and 3-sigma Gaussian post-filter were used for noise reduction in the EDS maps. 

\section*{Data Availability Statement}
The data supporting this study's findings are available from the corresponding author upon reasonable request.

\begin{acknowledgement}
M.X.~and J.M.L~acknowledge support for this work from the Air Force Office of Scientific Research (FA9550-20-0066) and the MIT Research Support Committee. We acknowledge support from the National Science Foundation (DMR-1751736). K.Y. acknowledges support from the National Science Foundation Graduate Research Fellowship, grant no. 1745302. This research was partly supported by a grant from the United States-Israel Binational Science Foundation (BSF), under grant no. 2020270. This research was supported in part by the Sagol Weizmann-MIT Bridge Program. This research was supported in part by the Skolkovo Institute of Science and Technology as part of the MIT-Skoltech Next Generation Program. This work made use of the MIT.nano Characterization Facilities.

\end{acknowledgement}

\begin{suppinfo}
% \section*{Supplementary Material}
See supplementary materials for additional film structure characterization, including in-plane/out-of-plane distances, nano-beam electron diffraction, and EDS maps. 
\end{suppinfo}

\bibliography{paperpile}
\end{document}